\documentclass[]{emulateapj}
\usepackage{apjfonts,epsfig}

\def\fsub{\mathrel{f_{\rm sub}}}
\def\tninety{\mathrel{t_{90}}}
\def\tfifty{\mathrel{t_{50}}}
\def\Ndm{\mathrel{N_{\rm DM}}}

\def\ls{\mathrel{\hbox{\rlap{\hbox{\lower4pt\hbox{$\sim$}}}\hbox{$<$}}}}
\def\gs{\mathrel{\hbox{\rlap{\hbox{\lower4pt\hbox{$\sim$}}}\hbox{$>$}}}}
\def\Mtot{\mathrel{M_{\rm tot}}}

\def\hkpc{\mathrel{h^{-1}{\rm kpc}}}

\lefthead{Smith \& Taylor}
\righthead{Cluster Substructure and Assembly Histories}

\slugcomment{Received 2008 April 25; Accepted 2008 June 24}

\begin{document}

\title{Connecting Substructure in Galaxy Cluster Cores at z=0.2 with
Cluster Assembly Histories}

\author{Graham P.\ Smith\altaffilmark{1,3} and 
        James E.\ Taylor\altaffilmark{2,3}}

\altaffiltext{1}{School of Physics and Astronomy, University of
                 Birmingham, Edgbaston, Birmingham, B15 2TT, England;
                 gps@star.sr.bham.ac.uk}  
\altaffiltext{2}{Department of Physics and Astronomy, University of
                 Waterloo, Waterloo, Ontario, N2L 3G1, Canada;
                 taylor@sciborg.uwaterloo.ca} 
\altaffiltext{3}{California Institute of Technology, Mail Code 105-24, 
                 1200 E.\ California Blvd., Pasadena, CA 91125}

\begin{abstract}
We use semi-analytic models of structure formation to interpret
gravitational lensing measurements of substructure in galaxy cluster
cores ($R{\le}250{\hkpc}$) at $z=0.2$.  The dynamic range of the
lensing-based substructure fraction measurements is well matched to
the theoretical predictions, both spanning $\fsub\sim0.05-0.65$.  The
structure formation model predicts that $\fsub$ is correlated with
cluster assembly history.  We use simple fitting formulae to
parameterize the predicted correlations:
$\Delta_{90}=\tau_{90}+\alpha_{90}\log(\fsub)$ and
$\Delta_{50}=\tau_{50}+\alpha_{50}\log(\fsub)$, where $\Delta_{90}$
and $\Delta_{50}$ are the predicted lookback times from $z=0.2$ to
when each theoretical cluster had acquired 90\% and 50\% respectively
of the mass it had at $z=0.2$.  The best-fit parameter values are:
$\alpha_{90}=(-1.34\pm0.79)\,{\rm Gyr}$,
$\tau_{90}=(0.31\pm0.56)\,{\rm Gyr}$ and
$\alpha_{50}=(-2.77\pm1.66)\,{\rm Gyr}$,
$\tau_{50}=(0.99\pm1.18)\,{\rm Gyr}$.  Therefore (i) observed clusters
with $\fsub\ls0.1$ (e.g.\ A\,383, A\,1835) are interpreted, on
average, to have formed at $z\gs0.8$ and to have suffered $\le10\%$
mass growth since $z\simeq0.4$, (ii) observed clusters with
$\fsub\gs0.4$ (e.g.\ A\,68, A\,773) are interpreted as, on average,
forming since $z\simeq0.4$ and suffering $>10\%$ mass growth in the
$\sim500\,{\rm Myr}$ preceding $z=0.2$, i.e.\ since $z=0.25$.  In
summary, observational measurements of $\fsub$ can be combined with
structure formation models to estimate the age and assembly history of
observed clusters.  The ability to ``age-date'' approximately clusters
in this way has numerous applications to the large clusters samples
that are becoming available.
\end{abstract}


\section{Introduction}\label{sec:intro}

The mass growth of clusters is sensitive to the dark energy equation
of state parameter $w$, the matter density of the universe $\Omega_M$
and the normalization of the matter power spectrum $\sigma_8$ (e.g.\
\citealt{Evrard93,Smith03,Mantz07}).  Clusters are inferred to grow
hierarchically via the ingestion of smaller dark matter halos (that
host galaxies) into the more massive parent halo (cluster).  The
structure of galaxy clusters, specifically the internal substructure
of clusters, therefore contains a wealth of cosmological information,
including possible clues about the physics of the dark matter particle
itself (e.g.\ \citealt{Natarajan02b}).  From an astrophysical point of
view, the mass growth of clusters brings new (generally gas rich)
galaxy populations into clusters (e.g.\ \citeauthor{Moran07}
\citeyear{Moran07}) and may lead to shock-heating of the intracluster
medium and/or disruption of cooling in cluster cores (e.g.\
\citealt{Poole08}).  Reliable measurement and intepretation of cluster
substructure is therefore of broad interest.

The most direct way to detect substructure within clusters is via
gravitational lensing.  Group-scale substructures within individual
clusters were detected in early ground-based strong-lensing studies of
individual clusters \citep{Pello91,Kneib93,Kneib95} and subsequently
measured to high precision using \emph{Hubble Space Telescope (HST)}
data \citep{Kneib96}.  \citeauthor{Smith05a} (\citeyear{Smith05a} --
hereafter Sm05 -- see \S\ref{sec:obs}) then measured the structure of
a sample of 10 clusters at $z{\simeq}0.2$.  In this article we use
\citeauthor{Taylor04}'s (\citeyear{Taylor04} -- hereafter TB04)
semi-analytical models of structure formation to interpret Sm05's
cluster substructure measurements, as a means of exploring
lensing-based substructure measurements as a quantitative probe of
cluster age and assembly history.

We summarize Sm05 and TB04 in \S\ref{sec:obs} and \S\ref{sec:theory}
respectively, and then synthesize observations and theory in
\S\ref{sec:results}.  We discuss caveats in \S\ref{sec:discuss} and
summarize our conclusions and discuss future prospects in
\S\ref{sec:conc}.  We assume $H_0{=}70{\rm km\,s^{-1}Mpc^{-1}}$,
${\Omega}_{\rm M}{=}0.3$, ${\Omega}_{\rm \Lambda}{=}0.7$ and
$\sigma_8=0.9$ throughout.  The lookback time from $z=0$ to $z=0.2$ is
$t_{z=0.2}=2.44\,{\rm Gyr}$ in this cosmology.

\section{Summary of Observational Results}\label{sec:obs}

\begin{table*}
\footnotesize
\caption{
Observational Substructure Measurements\label{tab:sample}
}
\begin{center}
\begin{tabular}{lccccllcc}
\hline
\hline
\noalign{\vspace{1mm}}
Cluster   & Redshift & ~~~~~~~~$\Mtot$ & ${\Ndm}$    & ${\fsub}$ & Strong Lensing? & Sm05 Classification & ~~~~~~~~${\Delta_{50}}$ & ${\Delta_{90}}$ \cr
          &          & ~~~~~~~~($10^{14}M_\odot$) &  &           &                 &                     & ~~~~~~~~$({\rm Gyr})$ & $({\rm Gyr})$ \cr
\noalign{\vspace{1mm}}
\hline
\noalign{\vspace{1mm}}
A\,383  & 0.188 & ~~~~~~~~$2.6\pm0.1$ & 1               & $0.06{\pm}0.01$ & Confirmed & Undisturbed uni-modal & ~~~~~~~~$4.3{\pm}2.4$ & $2.0{\pm}1.1$ \cr
A\,963  & 0.206 & ~~~~~~~~$2.4\pm0.2$ & 1               & $0.06{\pm}0.01$ & Confirmed & Undisturbed uni-modal & ~~~~~~~~$4.3{\pm}2.4$ & $2.0{\pm}1.1$ \cr
A\,1835 & 0.253 & ~~~~~~~~$4.1\pm1.1$ & 1               & $0.06{\pm}0.01$ & Unconfirmed & Undisturbed uni-modal & ~~~~~~~~$4.3{\pm}2.4$ & $2.0{\pm}1.1$ \cr
A\,267  & 0.230 & ~~~~~~~~$1.9\pm0.4$ & 1               & $0.08{\pm}0.01$ & Unconfirmed & Disturbed uni-modal & ~~~~~~~~$4.0{\pm}2.2$ & $1.8{\pm}1.0$ \cr
A\,1763 & 0.288 & ~~~~~~~~$1.5\pm0.8$ & 1               & $0.15{\pm}0.05$ & No? & Disturbed uni-modal & ~~~~~~~~$3.2{\pm}1.8$ & $1.4{\pm}0.9$ \cr
A\,209  & 0.209 & ~~~~~~~~$1.1\pm0.5$ & 1               & $0.18{\pm}0.06$ & No? & Disturbed uni-modal & ~~~~~~~~$3.0{\pm}1.7$ & $1.3{\pm}0.8$ \cr
A\,2219 & 0.228 & ~~~~~~~~$2.4\pm0.1$ & 3               & $0.20{\pm}0.01$ & Confirmed & Disturbed multi-modal & ~~~~~~~~$2.9{\pm}1.7$ & $1.3{\pm}0.8$ \cr
A\,2218 & 0.171 & ~~~~~~~~$4.0\pm0.1$ & 2               & $0.27{\pm}0.01$ & Confirmed & Disturbed multi-modal & ~~~~~~~~$2.5{\pm}1.5$ & $1.1{\pm}0.7$ \cr
A\,68   & 0.255 & ~~~~~~~~$3.1\pm0.1$ & 2               & $0.36{\pm}0.01$ & Confirmed & Disturbed multi-modal & ~~~~~~~~$2.2{\pm}1.4$ & $0.9{\pm}0.7$ \cr
A\,773  & 0.217 & ~~~~~~~~$3.6\pm1.2$ & 3               & $0.61{\pm}0.20$ & Unconfirmed & Disturbed multi-modal & ~~~~~~~~$1.6{\pm}1.3$ & $0.6{\pm}0.6$ \cr
\hline
\end{tabular}
\end{center}
\end{table*}

Sm05 investigated the projected mass and structure of ten X-ray
luminous ($L_X\ge4\times10^{44}\,{\rm erg\,s^{-1}}$, $0.1-2.4\,{\rm
keV}$) cluster cores at $0.17\le z\le0.25$
(Table~\ref{tab:sample}). \emph{Hubble Space Telescope (HST)}/WFPC2
imaging and ground-based spectroscopy of gravitational arcs
\citep{Smith01a,Smith02b,Sand05,Richard07}, were used to characterize
the strong and weak gravitational lensing signal of each cluster core.
The lensing signals were then used to constrain a detailed
parametrized model of the projected mass distribution in each cluster
core following \citeauthor{Kneib93Th} (\citeyear{Kneib93Th} -- see
also \citealt{Kneib96} and \citealt{Smith02Th}).  Each lens model
includes mass components that account for both the underlying dark
matter distribution in the cluster (cluster/group-scale mass
components) and the cluster galaxies down to
$L_K{\ge}0.1L^\star_K$. For clarity, we refer to the main central
cluster dark matter halo as the cluster-scale mass component, and all
other massive substructures associated with infallen clusters and/or
groups as group-scale mass components.  The typical number of
galaxy-scale mass components in each lens model was 30.

The \emph{HST} data probe out to a typical projected cluster-centric
radius of $R=250\hkpc$ at the cluster redshifts.  Sm05 measured the
cluster substructure within this radius.  Here we define $\fsub$ as the
fraction of mass associated with substructures, $\fsub{\equiv}M_{\rm
sub}/M_{\rm tot}=1-M_{\rm cen}/M_{\rm tot}$, where $M_{\rm cen}/M_{\rm
tot}$ is the central mass fraction used by Sm05.  For this purpose,
``substructures'' include group-scale masses and galaxy-scale masses, with
the exception of the brightest cluster galaxy (BCG).  This is because BCG
mass is degenerate with cluster-scale mass in the lens models, and, in any
event, Sm05 found no evidence for BCGs being offset from the center of the
cluster-scale mass components.  The clusters are listed in order of
increasing $\fsub$ in Table~\ref{tab:sample}, together with $\Ndm$, the
number of cluster/group-scale mass components in each lens model.  As
expected, $\Ndm$ and $\fsub$ are correlated, however there is a factor of
3 spread in $\fsub$ for sub-samples of clusters with $\Ndm=1$ and
$\Ndm>1$.  Note, that to aide comparison of Sm05's lens modeling results
with predictions, the values of $\fsub$ in Table~\ref{tab:sample} have
been adjusted upwards to take account of the fact that the ``L''-shaped
observed WFPC2 field of view ($\sim5\,{\rm arcmin}^2$) covers just
$\sim50\%$ of a circle of radius $250\hkpc$ ($\sim10\,{\rm arcmin}^2$).
The WFPC2 observations were originally designed to include the likely 
group-scale substructures in the cluster cores.  The adjustments to 
$\fsub$ therefore account statistically for the galaxy-scale masses not 
included in Sm05's analysis, leaving $\Ndm$ unchanged.

Cluster-to-cluster differences in $\fsub$ arise for two reasons: (i)
group-scale mass components in the lens models that are associated
with massive, likely in-falling structures such as groups of galaxies,
and (ii) cluster galaxies that are associated both with the central
cluster dark matter halo (and are presumably virialized) and with the
in-falling structures.  Sm05 broadly interpreted measurements of
$\fsub$ as indicating the merger history of clusters, however
quantitative conclusions on cluster assembly and age were impossible
without theoretical models.

\section{Summary of Theoretical Models}\label{sec:theory}

TB04's semi-analytic model of halo evolution provides a fast way of
generating a large number of model halos for comparison with
observations, and agrees well in its preditions with self-consistent
n-body simulations \citep{Taylor05b}.  The model includes two main
components, a merger-tree code for determining the assembly history of
an individual cluster, and an analytic description of how the main
halo and the merging subcomponent evolve after each merger. We will
summarize these components here, and refer the reader to TB04 for full
details.

The merger tree describing the assembly of a single cluster is
generated randomly using the algorithm of
\cite{Somerville99}. Starting from a halo of specified mass at $z=0$,
the algorithm chooses an interval to step back in redshift and picks
progenitors at this redshift following extended Press-Schechter merger
statistics \citep{LaceyCole}. Iterating produces a complete history of
the mergers through which the final object assembled, down to some
limiting mass resolution and back to some redshift. In this article we
use a set of 1000 merger trees whose final masses at $z=0$ are
randomly drawn from the massive end ($>5\times10^{14}M_\odot$) of a
halo mass function.  For the range of concentration parameters derived
for our halos (as explained below), this produces projected masses
within a cluster-centric radius of $R=250\hkpc$ (integrating out to
the virial radius along the line of sight) in the range
$0.4-8\times10^{14}M_{\odot}$, similar to those in the observed sample
(Table~\ref{tab:sample}). The merger trees for these systems are
followed back to a redshift of $50$, or until they drop below the
resolution limit of $10^{-4}\times$ the final mass.

Given a merger tree, the model of TB04 selects the most massive
progenitor in the most recent merger and traces its history back,
selecting the most massive progenitor at each time step. This object
is considered the ``main'' system, and is modelled as a spherical halo
with a radial density profile similar to the ``universal'' profile
found in simulations \citep{Navarro97}. To account for some of the
centrally concentrated (baryonic) mass in the cluster, we use the
fitting formula proposed by \cite{Moore98}, which has a steeper
$r^{-1.5}$ central cusp. In an update to the model presented in TB04,
in this work we use the whole mass assembly history of the system,
rather than just its instantaneous mass, to derive the value of its
concentration parameter at each redshift step, following
\cite{Wechsler02}.  Each merging subhalo is added to this main system
as a smaller spherical object with a Moore profile, and its subsequent
orbit and mass-loss history are determined using the analytic
description of dynamical friction, tidal stripping and tidal heating
in \cite{Taylor01}. Subhalos are tracked until they are disrupted
either by repeated tidal stripping or by passing within $0.01R_{vir}$
of the centre of the main potential. The model also includes simple
treatments of sub-substructure (subhalos within subhalos), correlated
orbits in infalling groups of objects, and collisions and encounters
between objects. See TB04 for a detailed explanation of these
elements.

For each merger tree, we use the semi-analytic model to determine how
much substructure exists at $z=0.2$, and calculate $\fsub$ as the 
fraction of the
mass projected within $250\hkpc$ of the halo centre that is contained
in the 30 most massive subhalos in the system (to match the
observations -- see \S\ref{sec:obs}).  We exclude from the final
analysis any substructure which ends up within 5\% of $R_{vir}$ of the
halo centre, since in a real cluster most such objects would interact
strongly and merge rapidly with the dominant central galaxy.  We
also record several indicators of the mass assembly history of the main
system in each merger tree, including $\tninety$ and $\tfifty$, the
lookback times by which it had assembled 90\% and 50\% respectively of
the mass it has at redshift $0.2$.  These thresholds are chosen to
probe the recent infall history and cluster age respectively. We show
the resulting $\fsub-\tninety$ and $\fsub-\tfifty$ distributions in
Fig.~\ref{fig:epoch}.  The observed clusters span a dynamic range of
$0.06{\le}{\fsub}{\le}0.62$ (Table~\ref{tab:sample}), which is
comparable with that of the TB04 theoretical models, and consistent
with other theoretical estimates of cluster substructure (e.g.\
\citealt{Diemand04,Gao04,Nagai05,vandenbosch05}).

\section{Results}\label{sec:results}

\begin{figure}
\centerline{
\psfig{file=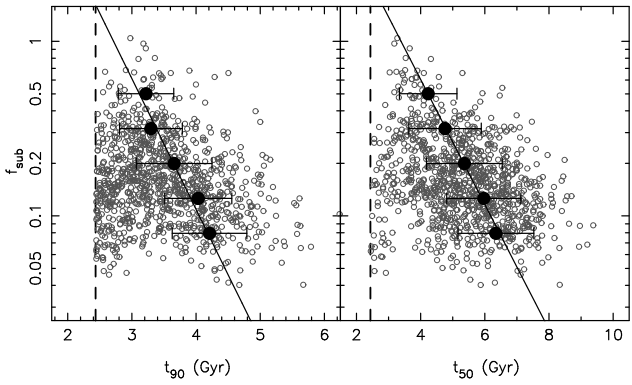,width=50mm,angle=-90}
}
\caption{Predicted distribution of substructure fraction versus
  $\tninety$ (left) and $\tfifty$ (right).  Note that the lookback
  time from the $z=0$ to $z=0.2$ is $2.4\,{\rm Gyr}$ (dashed line).
  The grey open points show clusters from TB04's model -- they span a
  similar dynamic range to the observed clusters (see
  Table~\ref{tab:sample}).  The big black points show the mean
  prediction in equally spaced logarithmic bins; the error bars show
  the $1\sigma$ scatter in each bin.  The solid black line shows the
  best fit relations described in the text (\S\ref{sec:results})
  \label{fig:epoch} }
\end{figure}

Fig.~\ref{fig:epoch} reveals that the TB04 model predicts that $\fsub$
and $\tninety$ are correlated, as are $\fsub$ and $\tfifty$.  However,
at $\fsub<0.2$ the distributions of ${\tninety}$ and $\tfifty$ are
bi-modal -- some theoretical clusters exhibiting both a low
substructure fraction and recent significant infall.  The unphysical
location of these clusters in Fig.~1 is due to a timing difference
between the axes -- $\tninety$ is sensitive to mass growth within the
virialized region of the cluster (approximately a sphere of diameter
$\sim3\,{\rm Mpc}$); $\fsub$ is sensitive to substructures in a
cylinder of diameter $500\hkpc$ through the center of the cluster.  If
an infalling structure has crossed the virial radius but not yet
entered the cylinder, then it may appear in the ``spike''of clusters
at $\fsub<0.2$ and $\tninety<3\,{\rm Gyr}$ (or $\tfifty<3\,{\rm
Gyr}$).  We exclude these clusters from the calculations described
below.

We have therefore confirmed and considerably extended Sm05's
qualitative interpretation of the lensing results: clusters with high
substructure fractions have (i) suffered more pronounced recent infall
than clusters with low substructure fractions, and (ii) formed at
later times than clusters with lower substructure fractions.  To
quantify this, we fit the following simple formulae to the theoretical
data:
$\Delta_{90}=\tninety-t_{z=0.2}=\tau_{90}+\alpha_{90}\log(\fsub)$ and
$\Delta_{50}=\tfifty-t_{z=0.2}=\tau_{50}+\alpha_{50}\log(\fsub)$,
where $\alpha_{90}$ and $\alpha_{50}$ parameterize the dependence of
$\Delta_{90}$ and $\Delta_{50}$ respectively on $\fsub$; $\tau_{90}$
and $\tau_{50}$ are the intercepts at $\fsub=1$.  We first bin up the
individual theoretical data points in bins of width
$\Delta\log(\fsub)=0.2$, and then weight each bin in the fit by the
reciprocal of the sample variance in the respective bin.  The best fit
parameter values obtained in this way are:
$\alpha_{90}=(-1.34\pm0.79)\,{\rm Gyr}$,
$\tau_{90}=(0.31\pm0.56)\,{\rm Gyr}$ and
$\alpha_{50}=(-2.77\pm1.66)\,{\rm Gyr}$,
$\tau_{50}=(0.99\pm1.18)\,{\rm Gyr}$.  The $\tau_{90}$ and $\tau_{50}$
parameters are measurements of the timing difference between the axes
in Fig.~\ref{fig:epoch} discussed above, and are well-matched to the
infall timescale of $\sim0.5-1\,{\rm Gyr}$.

The best-fit models were then used to interpret quantitatively the
observed substructure fractions from Sm05 listed in
Table~\ref{tab:sample} -- i.e.\ to estimate the age ($\Delta_{50}$)
and recent infall history ($\Delta_{90}$) of each cluster -- see
Table~\ref{tab:sample}.  The uncertainties on $\Delta_{90}$ and
$\Delta_{50}$ quoted in Table~\ref{tab:sample} incorporate errors on
$\fsub$ and on the best-fit parameter values derived above, and are
dominated by the scatter on $\fsub-\tfifty$ and $\fsub-\tninety$.  The
typical uncertainty on cluster age is $\sim1.7\,{\rm Gyr}$, with ages
spanning $\sim1-4\,{\rm Gyr}$ .  Cluster with the largest substructure
fractions, i.e.\ $\fsub\gs0.4$ formed within the $\sim2$Gyr preceding
$z=0.2$, i.e.\ since $z\ls0.4$, and had not assembled 90\% of the mass
they had at $z=0.2$ until $z\simeq0.25$, i.e.\ just $\sim500$Myr
before $z=0.2$.  In contrast, clusters with the lowest subtructure
fractions, i.e.\ $\fsub<0.1$ formed $\gs4\,{\rm Gyr}$ before $z=0.2$,
i.e.\ at $z\gs0.8$.  and then went on to assemble $>90\%$ of their
$z=0.2$ mass within the ensuing $\sim2\,{\rm Gyr}$, and suffered
negligible mass growth in the $\sim2\,{\rm Gyr}$ prior to $z=0.2$,
i.e.\ since $z\simeq0.4$.  Therefore we interpret clusters with the
lowest $\fsub$ as being, on average, almost fully ($\sim90\%$)
assembled by the time that the clusters with highest $\fsub$ had
barely formed (i.e.\ assembled half of their mass).

\section{Caveats}\label{sec:discuss}

TB04 make some simplifying assumptions that may affect our results:
(i) dark matter halos are assumed to be spherical, whereas real dark
matter halos are likely triaxial; (ii) all matter is treated as
collisonless, thus ignoring baryonic physics including adiabatic
contraction due to gas cooling and the dynamical effects of galaxies;
(iii) matter outside the cluster virial radius is ignored; in
contrast, lensing is sensitive to all the matter along the line of
sight, including chance foreground and background projections. We will
consider these complications in more detail in future work.

The main systematic uncertainty on the observed substructure fractions
is the completeness of Sm05's lens models as a function of sub-halo
mass.  A detailed study of this issue will be published in the future
(Hamilton-Morris et al., in prep.).  Here we identify how many of
Sm05's models may suffer from incompleteness.  Incompleteness is most
likely to arise from features in the dark matter distribution not
identified in the luminous properties of the clusters within
$R<250\hkpc$, including group-scale dark matter halos in which the
cluster galaxies may be embedded.  Strong lensing has been detected in
eight of the ten observed clusters (Table~\ref{tab:sample});
group-scale dark matter halos are detected in three of the eight by
the measurable way they alter the appearance of strongly-lensed arcs
compared to simpler mass distributions.  Clusters in which
strong-lensing has not been detected are therefore of greatest
concern, as pointed out in Sm05's discussion of the filament feeding
A\,1763; see also \citeauthor{Mercurio03}'s (\citeyear{Mercurio03})
discussion of the complex dynamical structure of A\,209.  Gross
substructure incompleteness may therefore be a concern for $\sim20\%$
of the observed sample.

\begin{figure}
\centerline{\psfig{file=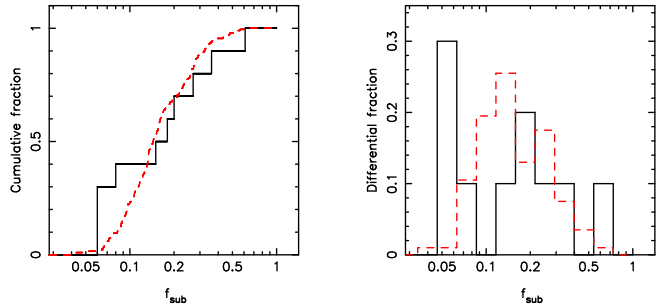,width=40mm,angle=-90}}
\caption{ Distribution of substructure fractions from the observed sample  
(black solid line) and from the theoretical model (red dashed).
The observed clusters show a possible excess of very low and very high
substructure systems -- this is discussed in \S\ref{sec:discuss}.  
\label{fig:fsub}}
\end{figure}

We also consider whether there is any evidence in the respective
$\fsub$ distributions for systematic differences between Sm05's X-ray
selected sample and TB04's mass-selected sample.  To address this we
compare the cumulative and differential $\fsub$ distributions in
Fig.~\ref{fig:fsub}.  A KS test obtains a probability of 20\% that the
observed and theoretical distributions are drawn from the same parent
distribution, i.e.\ formally evidence of systematic differences
between the samples at $1.3\sigma$ significance.  The small observed
sample size of just 10 clusters is clearly a limiting factor here.
The right panel of Fig.~\ref{fig:fsub} helps somewhat, as it
highlights more clearly the \emph{possible} bias of the observed
sample to very high and very low values of $\fsub$.  Such a bias would
be plausible because cluster-cluster mergers and cool cores both
suffer an excess of X-ray flux at a fixed mass over a non-merging,
non-cool core cluster.  A much larger observational sample is required
to investigate this issue further.

\section{Conclusions and Future Prospects}\label{sec:conc}

We have combined theoretical models of structure formation (TB04) with
gravitational lens models of galaxy clusters (Sm05) to explore how
measurements of cluster substructure from lensing observations can be
interpreted in the context of the age and assembly history of
clusters.  The main result is that $\fsub$, the fraction of cluster
mass within a projected cluster-centric radius of $R=250\hkpc$
associated with substructure (galaxies and group-scale halos), as can
be measured from lensing data, is predicted to be strongly correlated
with the age and recent mass growth of galaxy clusters.  We fitted the
the following simple formulae to the theoretical data to quantify the
predicted behavior in a convenient form:
$\Delta_{90}=\tninety-t_{z=0.2}=\tau_{90}+\alpha_{90}\log(\fsub)$ and
$\Delta_{50}=\tninety-t_{z=0.2}=\tau_{50}+\alpha_{50}\log(\fsub)$,
where $\tninety$ and $\tfifty$ are the lookback times at which a
cluster had acquired 90\% and 50\% of its mass at $z=0.2$.  The best
fit parameter values are: $\alpha_{90}=(-1.34\pm0.79)\,{\rm Gyr}$,
$\tau_{90}=(0.31\pm0.56)\,{\rm Gyr}$ and
$\alpha_{50}=(-2.77\pm1.66)\,{\rm Gyr}$,
$\tau_{50}=(0.99\pm1.18)\,{\rm Gyr}$.  Low-$\fsub$ clusters
($\fsub\ls0.1$; e.g.\ A\,383, A\,1835) are therefore interpreted as,
on average, having formed at $z{\gs}0.8$ and having suffered
${\le}10\%$ mass growth in the 2\,Gyr preceding $z=0.2$, i.e.\ since
$z\simeq0.4$.  In contrast, high-$\fsub$ clusters ($\fsub\gs0.4$;
e.g.\ A\,68, A\,773) are interpreted, on average, to have formed just
$\sim2\,{\rm Gyr}$ before $z=0.2$, i.e.\ since $z\simeq0.4$, and
suffered $10\%$ mass growth in the $\sim0.5\,{\rm Gyr}$ preceding
$z=0.2$, i.e.\ since $z{\simeq}0.25$.

Our synthesis therefore demonstrate that lensing-based measurements of
$\fsub$ can be combined with semi-analytic structure formation models
to estimate the average age and assembly history of observed clusters.
This suggests numerous avenues for further exploration, including: (i)
expansion of the observed samples by at least an order of magnitude,
(ii) calibration of the completeness of the lensing-based mass
function of sub-halos in clusters, (iii) investigation of how
lensing-based cluster age and assembly history estimates might allow
new cosmological constraints to be derived, for example, on the dark
energy equation of state parameter $w$, (iv) analysis of cluster
galaxy populations and cluster scaling relations as a function of
cluster age.

\acknowledgements
We thank Arif Babul, Habib Khosroshahi, Jean-Paul Kneib, Ian Smail and
Risa Wechsler for helpful discussions.  GPS acknowledges support from
NASA (HST-GO-10420.04-A), Caltech, and a Royal Society University
Research Fellowship.  JET acknowledges financial support from the US
NSF (grant AST-0307859) and DoE (contract DE-FG02-04ER41316) and from
NSERC Canada.


\begin{thebibliography}{31}
\expandafter\ifx\csname natexlab\endcsname\relax\def\natexlab#1{#1}\fi

\bibitem[{{Diemand} {et~al.}(2004){Diemand}, {Moore}, \& {Stadel}}]{Diemand04}
{Diemand}, J., {Moore}, B., \& {Stadel}, J. 2004, \mnras, 352, 535

\bibitem[{{Evrard} {et~al.}(1993){Evrard}, {Mohr}, {Fabricant}, \&
  {Geller}}]{Evrard93}
{Evrard}, A.~E., {Mohr}, J.~J., {Fabricant}, D.~G., \& {Geller}, M.~J. 1993,
  \apjl, 419, L9+

\bibitem[{{Gao} {et~al.}(2004){Gao}, {White}, {Jenkins}, {Stoehr}, \&
  {Springel}}]{Gao04}
{Gao}, L., {White}, S.~D.~M., {Jenkins}, A., {Stoehr}, F., \& {Springel}, V.
  2004, \mnras, 355, 819

\bibitem[{{Kneib}(1993)}]{Kneib93Th}
{Kneib}, J.-P. 1993, Ph.D.~Thesis, Universit\'e Paul Sabatier, Toulouse, France

\bibitem[{{Kneib} {et~al.}(1996){Kneib}, {Ellis}, {Smail}, {Couch}, \&
  {Sharples}}]{Kneib96}
{Kneib}, J.-P., {Ellis}, R.~S., {Smail}, I., {Couch}, W.~J., \& {Sharples},
  R.~M. 1996, \apj, 471, 643

\bibitem[{{Kneib} {et~al.}(1993){Kneib}, {Mellier}, {Fort}, \&
  {Mathez}}]{Kneib93}
{Kneib}, J.~P., {Mellier}, Y., {Fort}, B., \& {Mathez}, G. 1993, \aap, 273, 367

\bibitem[{{Kneib} {et~al.}(1995){Kneib}, {Mellier}, {Pello}, {Miralda-Escude},
  {Le Borgne}, {Boehringer}, \& {Picat}}]{Kneib95}
{Kneib}, J.~P., {Mellier}, Y., {Pello}, R., {Miralda-Escude}, J., {Le Borgne},
  J.-F., {Boehringer}, H., \& {Picat}, J.-P. 1995, \aap, 303, 27

\bibitem[{{Lacey} \& {Cole}(1993)}]{LaceyCole}
{Lacey}, C. \& {Cole}, S. 1993, \mnras, 262, 627

\bibitem[{{Mantz} {et~al.}(2007){Mantz}, {Allen}, {Ebeling}, \&
  {Rapetti}}]{Mantz07}
{Mantz}, A., {Allen}, S.~W., {Ebeling}, H., \& {Rapetti}, D. 2007, ArXiv
  e-prints, 709

\bibitem[{{Mercurio} {et~al.}(2003){Mercurio}, {Girardi}, {Boschin},
  {Merluzzi}, \& {Busarello}}]{Mercurio03}
{Mercurio}, A., {Girardi}, M., {Boschin}, W., {Merluzzi}, P., \& {Busarello},
  G. 2003, \aap, 397, 431

\bibitem[{{Moore} {et~al.}(1998){Moore}, {Governato}, {Quinn}, {Stadel}, \&
  {Lake}}]{Moore98}
{Moore}, B., {Governato}, F., {Quinn}, T., {Stadel}, J., \& {Lake}, G. 1998,
  \apjl, 499, L5+

\bibitem[{{Moran} {et~al.}(2007){Moran}, {Ellis}, {Treu}, {Smith}, {Rich}, \&
  {Smail}}]{Moran07}
{Moran}, S.~M., {Ellis}, R.~S., {Treu}, T., {Smith}, G.~P., {Rich}, R.~M., \&
  {Smail}, I. 2007, \apj, 671, 1503

\bibitem[{{Nagai} \& {Kravtsov}(2005)}]{Nagai05}
{Nagai}, D. \& {Kravtsov}, A.~V. 2005, \apj, 618, 557

\bibitem[{{Natarajan} {et~al.}(2002){Natarajan}, {Loeb}, {Kneib}, \&
  {Smail}}]{Natarajan02b}
{Natarajan}, P., {Loeb}, A., {Kneib}, J.-P., \& {Smail}, I. 2002, \apjl, 580,
  L17

\bibitem[{{Navarro} {et~al.}(1997){Navarro}, {Frenk}, \& {White}}]{Navarro97}
{Navarro}, J.~F., {Frenk}, C.~S., \& {White}, S.~D.~M. 1997, \apj, 490, 493

\bibitem[{{Pello} {et~al.}(1991){Pello}, {Sanahuja}, {Le Borgne}, {Soucail}, \&
  {Mellier}}]{Pello91}
{Pello}, R., {Sanahuja}, B., {Le Borgne}, J., {Soucail}, G., \& {Mellier}, Y.
  1991, \apj, 366, 405

\bibitem[{{Poole} {et~al.}(2008){Poole}, {Babul}, {McCarthy}, {Sanderson}, \&
  {Fardal}}]{Poole08}
{Poole}, G.~B., {Babul}, A., {McCarthy}, I.~G., {Sanderson}, A.~J.~R., \&
  {Fardal}, M.~A. 2008, ArXiv e-prints, 804

\bibitem[{{Richard} {et~al.}(2007){Richard}, {Kneib}, {Jullo}, {Covone},
  {Limousin}, {Ellis}, {Stark}, {Bundy}, {Czoske}, {Ebeling}, \&
  {Soucail}}]{Richard07}
{Richard}, J., {Kneib}, J.-P., {Jullo}, E., {Covone}, G., {Limousin}, M.,
  {Ellis}, R., {Stark}, D., {Bundy}, K., {Czoske}, O., {Ebeling}, H., \&
  {Soucail}, G. 2007, \apj, 662, 781

\bibitem[{{Sand} {et~al.}(2005){Sand}, {Treu}, {Ellis}, \& {Smith}}]{Sand05}
{Sand}, D.~J., {Treu}, T., {Ellis}, R.~S., \& {Smith}, G.~P. 2005, \apj, 627,
  32

\bibitem[{{Smith}(2002)}]{Smith02Th}
{Smith}, G.~P. 2002, Ph.D.~Thesis, University of Durham, UK

\bibitem[{{Smith} {et~al.}(2003){Smith}, {Edge}, {Eke}, {Nichol}, {Smail}, \&
  {Kneib}}]{Smith03}
{Smith}, G.~P., {Edge}, A.~C., {Eke}, V.~R., {Nichol}, R.~C., {Smail}, I., \&
  {Kneib}, J. 2003, \apjl, 590, L79

\bibitem[{{Smith} {et~al.}(2001){Smith}, {Kneib}, {Ebeling}, {Czoske}, \&
  {Smail}}]{Smith01a}
{Smith}, G.~P., {Kneib}, J., {Ebeling}, H., {Czoske}, O., \& {Smail}, I. 2001,
  \apj, 552, 493

\bibitem[{{Smith} {et~al.}(2005){Smith}, {Kneib}, {Smail}, {Mazzotta},
  {Ebeling}, \& {Czoske}}]{Smith05a}
{Smith}, G.~P., {Kneib}, J.-P., {Smail}, I., {Mazzotta}, P., {Ebeling}, H., \&
  {Czoske}, O. 2005, \mnras, 359, 417

\bibitem[{{Smith} {et~al.}(2002){Smith}, {Smail}, {Kneib}, {Davis}, {Takamiya},
  {Ebeling}, \& {Czoske}}]{Smith02b}
{Smith}, G.~P., {Smail}, I., {Kneib}, J.-P., {Davis}, C.~J., {Takamiya}, M.,
  {Ebeling}, H., \& {Czoske}, O. 2002, \mnras, 333, L16

\bibitem[{{Somerville} \& {Kolatt}(1999)}]{Somerville99}
{Somerville}, R.~S. \& {Kolatt}, T.~S. 1999, \mnras, 305, 1

\bibitem[{{Tasitsiomi} {et~al.}(2004){Tasitsiomi}, {Kravtsov}, {Gottl{\"o}ber},
  \& {Klypin}}]{Tasitsiomi04}
{Tasitsiomi}, A., {Kravtsov}, A.~V., {Gottl{\"o}ber}, S., \& {Klypin}, A.~A.
  2004, \apj, 607, 125

\bibitem[{{Taylor} \& {Babul}(2001)}]{Taylor01}
{Taylor}, J.~E. \& {Babul}, A. 2001, \apj, 559, 716

\bibitem[{{Taylor} \& {Babul}(2004)}]{Taylor04}
---. 2004, \mnras, 348, 811

\bibitem[{{Taylor} \& {Babul}(2005)}]{Taylor05b}
---. 2005, \mnras, 364, 535

\bibitem[{{van den Bosch} {et~al.}(2005){van den Bosch}, {Tormen}, \&
  {Giocoli}}]{vandenbosch05}
{van den Bosch}, F.~C., {Tormen}, G., \& {Giocoli}, C. 2005, \mnras, 359, 1029

\bibitem[{{Wechsler} {et~al.}(2002){Wechsler}, {Bullock}, {Primack},
  {Kravtsov}, \& {Dekel}}]{Wechsler02}
{Wechsler}, R.~H., {Bullock}, J.~S., {Primack}, J.~R., {Kravtsov}, A.~V., \&
  {Dekel}, A. 2002, \apj, 568, 52

\end{thebibliography}
\end{document}